\documentclass[aps,prd,final,twocolumn,floats,floatfix,nofootinbib,10pt]{revtex4-1}

\usepackage{blindtext}  
\usepackage{graphicx}
\usepackage{amsmath}
\usepackage{amsfonts}
\usepackage{amssymb}
\usepackage{amstext}
\usepackage[english]{babel}
\usepackage{enumitem}
\usepackage{helvet}
\usepackage{microtype}
\usepackage[pdftex]{hyperref}
\usepackage{multirow}
\usepackage{subcaption}

\input{Qcircuit.tex}

\begin{document}
\title{Quantum Circuits for Elementary Cellular Automata}

\author{Dmytro Fedoriaka}

\date{November 21, 2021}

\begin{abstract}
In this paper we identify full list of Elementary Cellular Automata rules which can be simulated using a quantum circuit (there are 22 such rules). For every such rule we present quantum circuit implementing it with $O(N)$ gates.
\end{abstract}

\maketitle

\section{Introduction}

Cellular Automata (CA) were first introduced by von Neumann \cite{neumann1966theory}, who wanted to find a model that was universal for computation and could in some sense replicate itself. Wolfram later defined Elementary Cellular Automata \cite{wolfram1983statistical} - simplest one-dimensional CA having 256 possible rules. All these rules are extensively researched \cite{voorhees1996computational, wolfram2002new, mcintosh2009one} and some of them display surprisingly complex behaviour.

Since Feynman \cite{feynman1982} introduced idea of quantum computer, scientists were suggesting different ideas for quantum analogs of classical computational models, including Cellular Automata.
\cite{grossing1988quantum,lent1993quantum,watrous1995one,meyer1996quantum} are some early works which define
quantum systems which are similar to classical CAs.  \cite{bleh2012quantum, arrighi2012quantum} suggest quantum analogs of famous 2-dimensional CA "Game of Life". Recent review \cite{Farrelly2020reviewofquantum} summarizes
modern research on Quantum Cellular Automata.

In that research, however, Quantum Cellular Automata are quantum systems, which only resemble classical CAs in some ways. In this paper we ask a question: can we construct a quantum system which would implement some classical CAs directly and exactly? We suggest to replace classical bits with qubits and find whether it's possible to apply unitary evolution on qubits, so it would transform basis states exactly as classical CA would transform classical bit arrays. 

We consider simplest classical CAs, namely Elementary CAs (ECAs). Not every ECA can be implemented by a unitary evolution, because such evolution can implement only reversible computation, and not all ECAs are reversible. However, some ECAs are reversible. In this paper we present circuits implementing all such ECAs. 

The paper is structured as follows. In \S 2 we formally define objective of this paper. In \S 3 we experimentally find all 22 reversible ECA rules. In \S 4 we describe how to build quantum circuits for all these 22 rules. 
In \S 5 we discuss our results and suggest directions of further research, and in \S 6 we conclude.

\section{Definitions and problem statement}

We define \textit{Elementary Cellular Automaton} (ECA) as an array of $N$ binary numbers: $a = [a_0, \dots, a_{N-1}] \in \{0, 1 \}^N$. A step of ECA is transformation of that array into other array $a^* = F(a)$ according to specified rule: $a_i^* = f(a_{i-1}, a_i, a_{i+1})$, where $a_i$ is value before transformation and $a^*_i$ - value after transformation.  

\textbf{Rules.} Function $f$ is a Boolean function of 3 variables and can be fully defined by table of values. If we read column of values in that table as a binary number, we will get 
$f(0,0,0)+2 \cdot f(0,0,1)+ 2^2 \cdot f(0,1,0) + 2^3 \cdot f(0,1,1) + 2^4 \cdot f(1,0,0) +  2^5 \cdot f(1,0,1)+ 2^6 \cdot f(1,1,0) + 2^7 \cdot f(1,1,1)$. This number is in range $[0, 255]$ and is called \textit{Wolfram code} 
(first defined in \cite{wolfram1983statistical}). This number is also referred to as "rule", as in "Rule 30". 

To fully define ECA, we also need to specify \textit{border conditions} (BCs), i.e. which cells are neighbors for $a_0$ and $a_{N-1}$. We will consider two BCs. At \textit{periodic BCs} we assume that $a_0$ and $a_{N-1}$ are neighbors. At \textit{fixed} BCs
we assume that tere are cells with value 0 to the left of $a_0$ and to the right of $a_{N-1}$, or, formally: $a^*_0 = f(0, a_0, a_1), a^*_{N-1}=f(a_{N-2}, a_{N-1}, 0)$.

To sum up, ECA is fully defined by rule $r \in [0, 255]$, number of cells $N$ and border condition.

\textbf{Quantum analog of ECA.} We consider array of qubits $q_0, \dots q_{N-1}$. We say that quantum circuit with unitary $U$ \textit{implements} certain ECA, if for every binary string $a \in \{0, 1 \}^N$, $U \ket{a} = \ket{a^*}$, where 
$\ket{a} = \ket{a_0, \dots a_{N-1}} = \ket{a_0} \otimes \dots \ket{a_{N-1}}$ is the basis state in which $q_i = \ket{0}$ if $a_i=0$ and $q_i=\ket{1}$ if $a_i=1$; and $a^*=F(a)$.

The objective of this paper is to find all ECA rules which can be implemented by a quantum circuit and for every such rule provide exact instruction how to construct circuit implementing it.

\section{Reversible rules}

\textbf{Lemma.} ECA with transition function $F: \{0, 1\}^N \to \{0, 1\}^N $ can be implemented by quantum circuit if and only if $F$ is injective.

\textbf{Proof.} If $F$ is injective, corresponding $U$ acts on basis states by permuting them. Therefore, $U$ is a permutation matrix. Permutation matrix is a unitary matrix, and any unitary matrix can be implemented by a quantum circuit \cite[\S 4.5]{nielsen2010quantum}.

If $F$ is not injective, then there are two different basis states $\ket{\psi_1}$ and $\ket{\psi_2}$ which $U$ must map to the same basis state. Therefore,  $U(\ket{\psi_1} - \ket{\psi_2})=0$, so $U$ is not unitary. In other words: there are two states mapped to the same state, so step of ECA is irreversible computation, and quantum circuits must be reversible. $\blacksquare$

We will call ECA \textit{reversible} if it transforms set of all $2^N$ states as injective function. According to the lemma above, reversible ECAs are exactly those ECAs which can be implemented by quantum circuit. To find all such ECAs, we conducted the following computational experiment.

For every rule $r \in [0, 255]$, for every $N \in [4, 20]$ and for both BCs, consider all possible states $a \in \{0, 1, \dots 2^N-1 \}$ (encoded as binary integers: $a \to \sum_i{2^i a_i}$). Then evaluate where ECA maps them after 1 step, and check if sequence $F(0), \dots F(2^N-1)$ is a permutation.

Results of this experiment (see \cite{repo}) showed that only 22 of 256 rules are reversible for some $N \in [4, 20]$ for some BCs. Namely:

\begin{itemize}[noitemsep,topsep=0pt]
\item Rules 51 and 204 - reversible always.
\item Rules 15, 85, 170, 240 - reversible for all $N$ at periodic BCs.
\item Rules 45, 75, 89, 101, 154, 166, 180, 210 - reversible for odd $N$ at periodic BCs.
\item Rules 105 and 150 - reversible at periodic BCs if $N \not \equiv 0 \pmod{3}$ 
      and at fixed BCs if $N \not \equiv 2 \pmod{3}$.
\item Rules 60, 102, 153, 195 - reversible for all $N$ at fixed BC.
\item Rules 90, 165 - reversible for even $N$ at fixed BC
\end{itemize}

In \S 4 we will build quantum circuits for these 22 rules.

\section{Circuits for rules}

First, for all reversible rules let's describe them by a compact formula rather then by table of values. For example, for Rule 60 formula is $a^*_i = f(a_{i-1}, a_i, a_{i+1}) = a_{i-1} \oplus a_i$. These formulas are given in Table \ref{table:rules}. We use the following notation: "$\oplus$" denotes addition modulo 2 (a.k.a. "XOR"), "$\cdot$" denotes multiplication (for Boolean variables equivalent to conjunction, or "AND"), $\overline{x} = x \oplus 1$ (equivalent to negation).

After analyzing these formulas we can see that there are only 5 non-trivial rules (170, 60, 90, 150, 166), and all other rules
are trivial or can be reduced to one of non-trivial rules. 

In what follows we describe circuits implementing all non-trivial circuits, then show how other rules can be reduced to them.
We only define how circuits act on basis states, so in this section we will talk of qubits as having Boolean values.
We use the following notation: $a_0, \dots a_{N-1}$ - initial binary values of qubits before circuit; $a'_0, \dots a'_{N-1}$ - values after intermediary steps; 
$a^*_0, \dots a^*_{N-1}$ - values after circuit. $q_0, \dots, q_{N-1}$ - qubits (as arguments to gates). 

We use the following gates: $SWAP$ - swap gate; $CNOT$ - controlled-X gate; $\overline{C}C..CNOT$ - multi-controlled gate which applies $X$ to last argument when first argument is 0 and all other arguments are 1. These gates (as well as their graphic representation used in this paper) are defined in \cite[\S 4]{nielsen2010quantum}.

\subsection{Rule 170}

Rule 170 has formula $a^*_i = a_{i+1}$. Therefore, at periodic BCs, it performs a 
circular shift to the left. To implement it, we need to shift qubits 
$q_1, \dots q_{N-1}$ to the left, and put $q_0$ at the rightmost position. 
This can be done by a sequence of $N-1$ SWAP gates: 
$SWAP(q_0,q_1); SWAP(q_1,q_2), \dots, SWAP(q_{N-2}, q_{N-1})$. Figure \ref{fig:example170}  shows example circuit for this rule for N=7.  

\subsection{Rule 60}

Rule 60 has formula $a^*_i = a_{i-1} \oplus a_i$. It XORs (in-place) every bit with it's left neighbor. This is exactly what CNOT does: it XORs (in-place) target qubit with control qubit. So, to implement Rule 60 (at fixed BCs), we need to apply $CNOT(q_{i}, q_{i+1})$ for $i=N-2, N-3, \dots 0$. Note that order matters because we must ensure that after qubit was modified, it's not used as a control. Figure \ref{fig:example60}  shows example circuit for this rule for N=7.

\subsection{Rule 90}

For this rule, as well as for rules 150 and 166 we will follow the following 3-step pattern to build the circuit. On step 1 we apply a series of local (2- or 3-qubit) gates from bottom to top to get $a'_i = a^*_{i-1}$ for $i \in [2, N-1]$. On step 2 we add some gates to get $a'_0 = a^*_{N-1}$ and $a'_1 = a^*_0$, without changing values of $a'_2, \dots a'_{N-1}$. On step 3 we apply left circular shift, getting desired final state.

Gates for step 1 will be obvious from the formula for the rule. Gates for step 2 will not be obvious and we would need to solve a system of Boolean equations to express $a^*_{0}, a^*_{N-1}$ as function of $a_{0}, a_{N-1}; a^*_1, a^*_2 , \dots, a^*_{N-2}$, and transform the solution to such a form which can be implemented with quantum circuit.

Going back to Rule 90, it has formula $a^*_i = a_{i-1} \oplus a_{i+1}$. It replaces every bit with XOR of its two neighbors. Step 1 for this rule must have effect $a'_i = a_{i-2} \oplus a_i$, which can be implemented by a CNOT gate.

Below are 3 steps to build circuit for Rule 90 at fixed border conditions for even N.

\textbf{Step 1.} $CNOT(q_{i}, q_{i+2})$ for $i=N-3,N-2,\dots,1,0$.

\textbf{Step 2.} $CNOT(q_i, q_0)$ for $i=N-2, N-4, \dots, 2, 0$.

\textbf{Step 3.} Left circular shift (see \S 4.1).

Figure \ref{fig:example90} shows example circuit for this rule for N=8.

\subsection{Rule 150}

Rule 150 has formula $a^*_i = a_{i-1} \oplus a_i \oplus a_{i+1}$. It XORs every bit with its left and right neighbor. Step 1 for this rule must have effect $a'_i = a_{i-2} \oplus a_{i-1} \oplus a_i$, which can be implemented by two CNOT gates.

Below are 3 steps to build circuit for Rule 150. There are 3 cases for step 2, depending on border conditions and the remainder of N modulo 3 (denoted here $N \% 3$).

\textbf{Step 1.} 

$CNOT(q_{i-2}, q_i); CNOT(q_{i-1}, q_i)$ for $i=N-1, N-2, \dots, 2$.

\textbf{Step 2} (periodic BCs, $N \% 3 = 1$).

$CNOT(q_i, q_0)$ for $i=(N-1)..1$ if $(N - i) \% 3 \ne 2$;

$CNOT(q_i, q_1)$ for $i=(N-1)..2$ if $(N - i) \% 3 \ne 0$.

\textbf{Step 2} (periodic BCs, $N \% 3 = 2$).

$SWAP(q_0, q_1)$;

$CNOT(q_i, q_0)$ for $i=(N-1)..2$ if $(N - i) \% 3 \ne 2$;   

$CNOT(q_i, q_1)$ for $i=(N-1)..2$ if $(N - i) \% 3 \ne 0$.

\textbf{Step 2} (fixed BCs, $N \% 3 \in \{0,1\}$).

$CNOT(q_0, q_1)$;

$CNOT(q_i, q_0)$ for $i=(N-1)..1$ if $(N - i) \% 3 \ne 2$.

\textbf{Step 3.} Left circular shift (see \S 4.1).

Figure \ref{fig:example150} shows example circuit for this rule for N=7 and periodic border conditions.
    
\subsection{Rule 166}

Rule 166 has formula $a^*_i = a_{i+1}\oplus(\overline{a_{i-1}}\cdot a_{i})$. Step 1 for this rule must have effect
$a'_i = a_{i}\oplus(\overline{a_{i-2}}\cdot a_{i-1})$. In other words, $a_i$ should be flipped if and only if 
$a_{i-1}=1$ and $a_{i-2}=0$. This can be achieved by $\overline{C}CNOT(q_{i-2}, q_{i-1}, q_i)$.

Below are 3 steps to build circuit for Rule 166 at periodic BC for odd N.

\textbf{Step 1.} $\overline{C}CNOT(q_{i-2}, q_{i-1}, q_i)$ for $i=(N-1)..2$.
        
\textbf{Step 2.}

$\overline{C}C..CNOT(q_{2i+2}; q_{2i+3}, q_{2i+5}, \dots, q_{N-2} ;q_0, q_1)$ \\
for $i=0,1,2,...,\lfloor N/2 \rfloor$.

$\overline{C}C..CNOT(q_{2i+1}; q_{2i+2}, q_{2i+4}, \dots, q_{N-3}, q_{N-1} ;q_0)$ \\
for $i=0,1,2,...,\lfloor N/2 \rfloor$.

\textbf{Step 3.} Left circular shift (see \S 4.1).

Figure \ref{fig:example166} shows example circuit for this rule for N=9.

\subsection{Other rules}

Below we give instructions how to build circuits for other 17 rules. We denote rule number by $r$ and unitary for corresponding circuit $U_r$.

Rules 15, 45, 51, 75, 85, 89, 101, 105, 153, 165, 195 have "$\oplus 1$" in their formula. This means they apply some other rule and then flip state of every cell. So they can be implemented by circuit for that other rule followed by X gate on every qubit. This can be formally written as $U_r = X^{\otimes N} \cdot U_{255-r}$.

Formulas for rules 102, 180, 210 and 240 can be obtained, correspondingly, from rules  60, 166, 154 and 170 by swapping $a_{i-1}$ and $a_{i+1}$. This means they are "reverses" or "mirror reflections". To implement them, we can reverse all qubits, implement the other rule, and reverse qubits again. Or, we can just vertically reflect graphical representation of the other circuit - to the same effect. Let's call this operation on circuit $Rev$, then we can formally write $U_{102} = Rev(U_{60})$ and so on.

Rule 154 can be implemented by flipping all bits, applying Rule 166 and flipping all bits again. Indeed,
$\overline{\overline{a_{i+1}} \oplus (\overline{\overline{a_i}}\cdot \overline{a_{i-1}})}
=a_{i+1} \oplus (a_i \cdot \overline{a_{i-1}}) $. Therefore, circuit for Rule 154 is: X gate on every qubit, then circuit for Rule 166 and then again X gate on every qubit. Formally, $U_{154} = X^{\otimes N} \cdot U_{166} \cdot X^{\otimes N}$.

Rule 204 doesn't change the state, so it corresponds to empty circuit (with matrix $I^{\otimes N}$, where I is the identity matrix).

With this, we have given explicit instruction how to build a circuit for all 22 reversible rules for all values of $N \ge 3$ and border conditions when these rules are reversible. These results are summarized in last column in Table \ref{table:rules}.

\section{Discussion}

\subsection{Proofs of irreversibility} 

This paper claims that 233 of 255 rules are irreversible for both BCs for any $N > 3$, and therefore can't be implemented by a quantum circuit. We checked that numerically for all $N \le 20$. We don't give rigorous proof for that statement, because reversibility of ECAs have been thoroughly studied in literature (e.g. \cite[\S 12]{mcintosh2009one}).

However, one way to rigorously prove that certain ECA is irreversible is to provide two different configurations which after one step are transformed to the same configuration. We wrote a program that finds such proofs for $N=6$ for all irreversible ECAs (can be found at \cite{repo}). One has to look at these proofs and generalize them to arbitrary $N$.

As an example, we will give such proofs of irreversibility for cases which we find interesting, i.e. for ECAs whose reversibility depends on remainder of N modulo 2 or 3. In examples below $s^k$ stands for sequence $s$ repeated $k$ times. 

\begin{itemize}[noitemsep,topsep=0pt]
    \item Rule 90 at fixed BCs is irreversible for odd N. 
          Proof: both states $[0]^{2k+1}$
          and $[1,0]^k,1$ are mapped on the next step to $[0]^{2k+1}$.
    \item Rule 150 at periodic BCs is irreversible if $N \equiv 0 \pmod{3}$. 
          Proof: both states $[0]^{3k}$
          and $[1,1,0]^k$ are mapped on the next step to $[0]^{3k}$.
    \item Rule 150 at fixed BCs is irreversible if $N \equiv 2 \pmod{3}$. 
          Proof: both states $[0]^{3k+2}$
          and $[1, 1, 0]^k, 1, 1$ are mapped on the next step to $[0]^{3k+2}$.
    \item Rule 166 at periodic BCs is irreversible for even N. 
          Proof: both states $[1,0]^k$
          and $[0, 1]^k$ are mapped on the next step to $[1,1]^k$.
          
\end{itemize}

\subsection{Complexity of circuits}

By construction all presented circuits have $O(N)$ gates. All circuits except for Rule 166 (and 7 rules reduced to it) consist of gates acting on at most 3 qubits. Circuit for Rule 166 has gates acting on up to $\lfloor N/2 \rfloor + 2 $
gates.

\subsection{Further research}

We identify the following directions in which research done in this paper can be continued.

1. \textbf{Fractional steps.} For every rule $r$ try to find corresponding Hamiltonian $H_r$ such that $U_r = \exp(-\frac{i H_r \Delta t}{\hbar})$ ($\Delta t$ - some fixed amount of time). Then if we let the system evolve under this Hamiltonian for time $k \Delta t$ ($k \in \mathbb{Z}$), it would implement $k$ steps of the ECA. If we now take fractional $k$, we would get evolution of ECA for fractional number of steps -- something which we can't do classically.

2. Study how these circuits act on non-basis states.

3. Study how we can implement reversible rules if we allow ancillary qubits and measurements.

4. Implement presented circuits on real quantum computers.

5. For Rule 166 find circuit with $O(N)$ gates each acting on $O(1)$ qubits, or prove that it's impossible.

6. Study if it's possible to get time complexity better than $O(N)$ by applying gates are applied in parallel. We might need to use additional qubits for that.

7. Find circuits for realistic physical constraints. For example, if we allow only nearest-neighbor coupling.

8. Find circuits for other (not elementary) CAs.

\section{Conclusion}

In this paper we have found full list of 22 ECA rules which can be implemented with a quantum circuit. In Table \ref{table:rules} we summarize this result: for every rule we present conditions under which it can be implemented by a quantum circuit and instruction how to build a circuit for it. All circuits have $O(N)$ gates, and 14 of them have gates acting on at most 3 qubits.

We also present a general technique for implementing ECAs with quantum circuits.

We supplement this paper with a GitHub repository \cite{repo}, which contains: Python code implementing all possible rules as quantum circuits (using Cirq library \cite{cirq}); tests for that code verifying their correctness for small values of $N$; code and results for experiment which we used to find all reversible rules and reversibility conditions in 
$\S 2$; code used to generate proofs of irreversibility (\S 5.1); simulation of circuits for some ECAs.

\begin{table*}[t]
\caption{All ECA rules that can be implemented with a quantum circuit.}
\label{table:rules}
\begin{tabular}{|l|l|ll|l|}
\hline
\multicolumn{1}{|c|}{\multirow{2}{*}{Rule}} & \multicolumn{1}{c|}{\multirow{2}{*}{Formula for $a^*_i$}}   & \multicolumn{2}{c|}{Reversible?}                                 & \multicolumn{1}{c|}{\multirow{2}{*}{Circuit ($U_r$)}} \\ \cline{3-4}
\multicolumn{1}{|c|}{}                      & \multicolumn{1}{c|}{}                                     & \multicolumn{1}{c|}{Periodic BC} & \multicolumn{1}{c|}{Fixed BC} & \multicolumn{1}{c|}{}                         \\ \hline
15                                          & $a_{i-1}+1$                                               & \multicolumn{1}{l|}{Yes}         & No                            & $X^{\otimes N} \cdot U_{240}$                                   \\ \hline
45                                          & $a_{i-1} \oplus (\overline{a_i} \cdot a_{i+1}) \oplus 1$ & \multicolumn{1}{l|}{For odd N}   & No                            & $X^{\otimes N} \cdot U_{210}$                                   \\ \hline
51                                          & $a_i \oplus 1$                                            & \multicolumn{1}{l|}{Yes}         & Yes                           & $X^{\otimes N}$                       \\ \hline
60                                          & $a_{i-1} \oplus a_i$                                      & \multicolumn{1}{l|}{No}          & Yes                           & See \S 4.2                     \\ \hline
75                                          & $a_{i-1}\oplus(\overline{a_{i+1}}\cdot a_i)\oplus 1$  & \multicolumn{1}{l|}{For odd N}   & No                            & $X^{\otimes N} \cdot U_{180}$                                  \\ \hline
85                                          & $a_{i+1}\oplus 1$  & \multicolumn{1}{l|}{Yes}   & No                            & $X^{\otimes N} \cdot U_{170}$                                   \\ \hline
89                                          & $a_{i+1}\oplus(\overline{a_{i-1}}\cdot a_{i}) \oplus 1$   & \multicolumn{1}{l|}{For odd N}   & No                            & $X^{\otimes N} \cdot U_{166}$     \\ \hline
90                                          & $a_{i-1} \oplus a_{i+1}$                                  & \multicolumn{1}{l|}{No}          & For even N                    & See \S 4.3                                    \\ \hline
101                                         & $a_{i+1}\oplus(\overline{a_{i}}\cdot a_{i-1}) \oplus 1$   & \multicolumn{1}{l|}{For odd N}   & No                            & $X^{\otimes N} \cdot U_{154}$                               \\ \hline
102                                         & $a_i \oplus a_{i+1}$                                      & \multicolumn{1}{l|}{No}          & Yes                           & $\text{Rev}(U_{160})$              \\ \hline
105                                         & $a_{i-1} \oplus a_i \oplus a_{i+1} \oplus 1$              & \multicolumn{1}{l|}{$N \% 3 \ne 0$} & $N \% 3 \ne 2$                              &  $X^{\otimes N} \cdot U_{150}$                                 \\ \hline
150                                         & $a_{i-1} \oplus a_i \oplus a_{i+1}$                       & \multicolumn{1}{l|}{$N \% 3 \ne 0$}            & $N \% 3 \ne 2$                              &    See \S 4.4                                  \\ \hline
153                                         & $a_i \oplus a_{i+1} \oplus 1$                             & \multicolumn{1}{l|}{No}          & Yes                           &  $X^{\otimes N} \cdot U_{102}$      \\ \hline
154                                         & $a_{i+1}\oplus(\overline{a_{i}}\cdot a_{i-1})$            & \multicolumn{1}{l|}{For odd N}   & No                            &  $ X^{\otimes N} \cdot U_{166} \cdot X^{\otimes N}$    \\ \hline
165                                         & $a_{i-1}\oplus a_{i+1} \oplus 1$                          & \multicolumn{1}{l|}{No}          & For even N                    &  $X^{\otimes N} \cdot U_{90} $                                 \\ \hline
166                                         & $a_{i+1}\oplus(\overline{a_{i-1}}\cdot a_{i})$            & \multicolumn{1}{l|}{For odd N}   & No                            &  See \S 4.5                                 \\ \hline
170                                         & $a_{i+1}$                                                 & \multicolumn{1}{l|}{Yes}         & No                            &  See \S 4.1                                   \\ \hline
180                                         & $a_{i-1}\oplus(\overline{a_{i+1}}\cdot a_i)$              & \multicolumn{1}{l|}{For odd N}   & No                            &  $\text{Rev}(U_{166})$                \\ \hline
195                                         & $a_{i-1}\oplus a_i \oplus a_{i+1}$                        & \multicolumn{1}{l|}{No}          & Yes                            & $X^{\otimes N} \cdot U_{60}$                                  \\ \hline
204                                         & $a_i$                                                     & \multicolumn{1}{l|}{Yes}         & Yes                            & $  I^{\otimes N}$                     \\ \hline
210                                         & $a_{i-1}\oplus(\overline{a_i}\cdot a_{i+1})$              & \multicolumn{1}{l|}{For odd N}   & No                             & $  \text{Rev}(U_{154})$                 \\ \hline
240                                         &  $a_{i-1}$                                                & \multicolumn{1}{l|}{Yes}         & No                             & $\text{Rev}(U_{170}) = U_{170}^\dag$                 \\ \hline
\end{tabular}
\end{table*}

\begin{figure*}
    \caption{Example circuits for Elementary Cellular Automata}
    \label{fig:fig1}
    \centering
    \begin{subfigure}[c]{0.18\textwidth}
        \[
        \Qcircuit @R=1.5em @C=0.7em {
         \\
         &\lstick{\text{0}}& \qw&\qswap \qw& \qw& \qw& \qw& \qw& \qw&\qw\\
         &\lstick{\text{1}}& \qw&\qswap\qwx  \qw&\qswap \qw& \qw& \qw& \qw& \qw&\qw\\
         &\lstick{\text{2}}& \qw& \qw&\qswap\qwx  \qw&\qswap \qw& \qw& \qw& \qw&\qw\\
         &\lstick{\text{3}}& \qw& \qw& \qw&\qswap\qwx  \qw&\qswap \qw& \qw& \qw&\qw\\
         &\lstick{\text{4}}& \qw& \qw& \qw& \qw&\qswap\qwx  \qw&\qswap \qw& \qw&\qw\\
         &\lstick{\text{5}}& \qw& \qw& \qw& \qw& \qw&\qswap\qwx  \qw&\qswap \qw&\qw\\
         &\lstick{\text{6}}& \qw& \qw& \qw& \qw& \qw& \qw&\qswap\qwx  \qw&\qw\\
         \\
        }
        \]
        \caption{Rule 170, N=7, periodic BC}
        \label{fig:example170}
    \end{subfigure}
    \begin{subfigure}[c]{0.30\textwidth}
        \[
        \Qcircuit @R=1em @C=0.7em {
         \\
         &\lstick{\text{0}}& \qw& \qw  & \qw  & \qw  & \qw  & \qw  &\control \qw  &\qw\\
         &\lstick{\text{1}}& \qw& \qw  & \qw  & \qw  & \qw  &\control \qw  &\targ  \qw\qwx&\qw\\
         &\lstick{\text{2}}& \qw& \qw  & \qw  & \qw  &\control \qw  &\targ  \qw\qwx& \qw  &\qw\\
         &\lstick{\text{3}}& \qw& \qw  & \qw  &\control \qw  &\targ  \qw\qwx& \qw  & \qw  &\qw\\
         &\lstick{\text{4}}& \qw& \qw  &\control \qw  &\targ  \qw\qwx& \qw  & \qw  & \qw  &\qw\\
         &\lstick{\text{5}}& \qw&\control \qw  &\targ  \qw\qwx& \qw  & \qw  & \qw  & \qw  &\qw\\
         &\lstick{\text{6}}& \qw&\targ  \qw\qwx& \qw  & \qw  & \qw  & \qw  & \qw  &\qw\\
         \\
        }
        \]
     \caption{Rule 60, N=7, fixed BC}
     \label{fig:example60}
    \end{subfigure}
    \begin{subfigure}[c]{0.50\textwidth}
        \[
        \Qcircuit @R=1em @C=0.7em {
         \\
         &\lstick{\text{0}}& \qw& \qw  & \qw  & \qw  & \qw  & \qw  &\control \qw  &\targ  \qw  &\targ  \qw  &\targ  \qw  &\qswap \qw& \qw& \qw& \qw& \qw& \qw& \qw&\qw\\
         &\lstick{\text{1}}& \qw& \qw  & \qw  & \qw  & \qw  &\control \qw  & \qw\qwx& \qw\qwx& \qw\qwx& \qw\qwx&\qswap\qwx  \qw&\qswap \qw& \qw& \qw& \qw& \qw& \qw&\qw\\
         &\lstick{\text{2}}& \qw& \qw  & \qw  & \qw  &\control \qw  & \qw\qwx&\targ  \qw\qwx& \qw\qwx& \qw\qwx&\control \qw\qwx& \qw&\qswap\qwx  \qw&\qswap \qw& \qw& \qw& \qw& \qw&\qw\\
         &\lstick{\text{3}}& \qw& \qw  & \qw  &\control \qw  & \qw\qwx&\targ  \qw\qwx& \qw  & \qw\qwx& \qw\qwx& \qw  & \qw& \qw&\qswap\qwx  \qw&\qswap \qw& \qw& \qw& \qw&\qw\\
         &\lstick{\text{4}}& \qw& \qw  &\control \qw  & \qw\qwx&\targ  \qw\qwx& \qw  & \qw  & \qw\qwx&\control \qw\qwx& \qw  & \qw& \qw& \qw&\qswap\qwx  \qw&\qswap \qw& \qw& \qw&\qw\\
         &\lstick{\text{5}}& \qw&\control \qw  & \qw\qwx&\targ  \qw\qwx& \qw  & \qw  & \qw  & \qw\qwx& \qw  & \qw  & \qw& \qw& \qw& \qw&\qswap\qwx  \qw&\qswap \qw& \qw&\qw\\
         &\lstick{\text{6}}& \qw& \qw\qwx&\targ  \qw\qwx& \qw  & \qw  & \qw  & \qw  &\control \qw\qwx& \qw  & \qw  & \qw& \qw& \qw& \qw& \qw&\qswap\qwx  \qw&\qswap \qw&\qw\\
         &\lstick{\text{7}}& \qw&\targ  \qw\qwx& \qw  & \qw  & \qw  & \qw  & \qw  & \qw  & \qw  & \qw  & \qw& \qw& \qw& \qw& \qw& \qw&\qswap\qwx  \qw&\qw\\
         \\
        }
        \]
         \caption{Rule 90, N=8, fixed BC}
         \label{fig:example90}
    \end{subfigure}
    \begin{subfigure}[c]{1\textwidth}
        \[
        \Qcircuit @R=1em @C=0.75em {
         \\
         &\lstick{\text{0}}& \qw& \qw  & \qw  & \qw  & \qw  & \qw  & \qw  & \qw  & \qw  &\control \qw  & \qw  &\targ  \qw  &\targ  \qw  &\targ  \qw  &\targ  \qw  & \qw  & \qw  & \qw  & \qw  &\qswap \qw& \qw& \qw& \qw& \qw& \qw&\qw\\
         &\lstick{\text{1}}& \qw& \qw  & \qw  & \qw  & \qw  & \qw  & \qw  &\control \qw  & \qw  & \qw\qwx&\control \qw  & \qw\qwx& \qw\qwx& \qw\qwx&\control \qw\qwx&\targ  \qw  &\targ  \qw  &\targ  \qw  &\targ  \qw  &\qswap\qwx  \qw&\qswap \qw& \qw& \qw& \qw& \qw&\qw\\
         &\lstick{\text{2}}& \qw& \qw  & \qw  & \qw  & \qw  &\control \qw  & \qw  & \qw\qwx&\control \qw  &\targ  \qw\qwx&\targ  \qw\qwx& \qw\qwx& \qw\qwx& \qw\qwx& \qw  & \qw\qwx& \qw\qwx& \qw\qwx&\control \qw\qwx& \qw&\qswap\qwx  \qw&\qswap \qw& \qw& \qw& \qw&\qw\\
         &\lstick{\text{3}}& \qw& \qw  & \qw  &\control \qw  & \qw  & \qw\qwx&\control \qw  &\targ  \qw\qwx&\targ  \qw\qwx& \qw  & \qw  & \qw\qwx& \qw\qwx&\control \qw\qwx& \qw  & \qw\qwx& \qw\qwx&\control \qw\qwx& \qw  & \qw& \qw&\qswap\qwx  \qw&\qswap \qw& \qw& \qw&\qw\\
         &\lstick{\text{4}}& \qw&\control \qw  & \qw  & \qw\qwx&\control \qw  &\targ  \qw\qwx&\targ  \qw\qwx& \qw  & \qw  & \qw  & \qw  & \qw\qwx&\control \qw\qwx& \qw  & \qw  & \qw\qwx& \qw\qwx& \qw  & \qw  & \qw& \qw& \qw&\qswap\qwx  \qw&\qswap \qw& \qw&\qw\\
         &\lstick{\text{5}}& \qw& \qw\qwx&\control \qw  &\targ  \qw\qwx&\targ  \qw\qwx& \qw  & \qw  & \qw  & \qw  & \qw  & \qw  & \qw\qwx& \qw  & \qw  & \qw  & \qw\qwx&\control \qw\qwx& \qw  & \qw  & \qw& \qw& \qw& \qw&\qswap\qwx  \qw&\qswap \qw&\qw\\
         &\lstick{\text{6}}& \qw&\targ  \qw\qwx&\targ  \qw\qwx& \qw  & \qw  & \qw  & \qw  & \qw  & \qw  & \qw  & \qw  &\control \qw\qwx& \qw  & \qw  & \qw  &\control \qw\qwx& \qw  & \qw  & \qw  & \qw& \qw& \qw& \qw& \qw&\qswap\qwx  \qw&\qw\\
         \\
        }
        \]
        \caption{Rule 150, N=7, periodic BC}
        \label{fig:example150}
    \end{subfigure}
        \begin{subfigure}[c]{1\textwidth}
        \[
        \Qcircuit @R=1em @C=0.75em {
         \\
         &\lstick{\text{0}}& \qw&  \qw  &  \qw  &  \qw  &  \qw  &  \qw  &  \qw  &\ctrlo{0} \qw  &\control \qw  &\control \qw  &\control \qw  &\control \qw  &\targ \qw  &\targ \qw  &\targ \qw  &\targ \qw  &\qswap \qw& \qw& \qw& \qw& \qw& \qw& \qw& \qw&\qw\\
         &\lstick{\text{1}}& \qw&  \qw  &  \qw  &  \qw  &  \qw  &  \qw  &\ctrlo{0} \qw  &\control \qw\qwx&\targ \qw\qwx&\targ \qw\qwx&\targ \qw\qwx&\targ \qw\qwx&\ctrlo{0} \qw\qwx&  \qw\qwx&  \qw\qwx&  \qw\qwx&\qswap\qwx  \qw&\qswap \qw& \qw& \qw& \qw& \qw& \qw& \qw&\qw\\
         &\lstick{\text{2}}& \qw&  \qw  &  \qw  &  \qw  &  \qw  &\ctrlo{0} \qw  &\control \qw\qwx&\targ \qw\qwx&\ctrlo{0} \qw\qwx&  \qw\qwx&  \qw\qwx&  \qw\qwx&\control \qw\qwx&  \qw\qwx&  \qw\qwx&  \qw\qwx& \qw&\qswap\qwx  \qw&\qswap \qw& \qw& \qw& \qw& \qw& \qw&\qw\\
         &\lstick{\text{3}}& \qw&  \qw  &  \qw  &  \qw  &\ctrlo{0} \qw  &\control \qw\qwx&\targ \qw\qwx&  \qw  &\control \qw\qwx&  \qw\qwx&  \qw\qwx&  \qw\qwx&  \qw\qwx&\ctrlo{0} \qw\qwx&  \qw\qwx&  \qw\qwx& \qw& \qw&\qswap\qwx  \qw&\qswap \qw& \qw& \qw& \qw& \qw&\qw\\
         &\lstick{\text{4}}& \qw&  \qw  &  \qw  &\ctrlo{0} \qw  &\control \qw\qwx&\targ \qw\qwx&  \qw  &  \qw  &  \qw\qwx&\ctrlo{0} \qw\qwx&  \qw\qwx&  \qw\qwx&\control \qw\qwx&\control \qw\qwx&  \qw\qwx&  \qw\qwx& \qw& \qw& \qw&\qswap\qwx  \qw&\qswap \qw& \qw& \qw& \qw&\qw\\
         &\lstick{\text{5}}& \qw&  \qw  &\ctrlo{0} \qw  &\control \qw\qwx&\targ \qw\qwx&  \qw  &  \qw  &  \qw  &\control \qw\qwx&\control \qw\qwx&  \qw\qwx&  \qw\qwx&  \qw\qwx&  \qw\qwx&\ctrlo{0} \qw\qwx&  \qw\qwx& \qw& \qw& \qw& \qw&\qswap\qwx  \qw&\qswap \qw& \qw& \qw&\qw\\
         &\lstick{\text{6}}& \qw&\ctrlo{0} \qw  &\control \qw\qwx&\targ \qw\qwx&  \qw  &  \qw  &  \qw  &  \qw  &  \qw\qwx&  \qw\qwx&\ctrlo{0} \qw\qwx&  \qw\qwx&\control \qw\qwx&\control \qw\qwx&\control \qw\qwx&  \qw\qwx& \qw& \qw& \qw& \qw& \qw&\qswap\qwx  \qw&\qswap \qw& \qw&\qw\\
         &\lstick{\text{7}}& \qw&\control \qw\qwx&\targ \qw\qwx&  \qw  &  \qw  &  \qw  &  \qw  &  \qw  &\control \qw\qwx&\control \qw\qwx&\control \qw\qwx&  \qw\qwx&  \qw\qwx&  \qw\qwx&  \qw\qwx&\ctrlo{0} \qw\qwx& \qw& \qw& \qw& \qw& \qw& \qw&\qswap\qwx  \qw&\qswap \qw&\qw\\
         &\lstick{\text{8}}& \qw&\targ \qw\qwx&  \qw  &  \qw  &  \qw  &  \qw  &  \qw  &  \qw  &  \qw  &  \qw  &  \qw  &\ctrlo{0} \qw\qwx&\control \qw\qwx&\control \qw\qwx&\control \qw\qwx&\control \qw\qwx& \qw& \qw& \qw& \qw& \qw& \qw& \qw&\qswap\qwx  \qw&\qw\\
         \\
        }
        \]
        \caption{Rule 166, N=9, periodic BC}
        \label{fig:example166} 
    \end{subfigure}
\end{figure*}


\bibliographystyle{utphys}    
\bibliography{references}     

\end{document}